\input harvmac
\input psfig.sty
\writedefs
%\draftmode

\def\vp{{\bf p}}

\def \ep{\epsilon}

\def\H {{\cal H}}
\def \B {{\cal B}}

\def \k {\kappa} 
\def \F {{\cal F}}
\def \g {\gamma}
\def \del {\partial}
\def \bd {\bar \partial }

\def \a {\alpha}
\def \b {\beta}
\def \chi {\chi}\def\r {\rho}
\def \s {\sigma}
\def \p {\phi}
\def \m {\mu}
\def \n {\nu}
\def \vp {\varphi }
\def \l {\lambda}

\def \td {\tilde }

\def \inv {^{-1}}
\def \ov {\over }
\def \four{{\textstyle{1\over 4}}}

\def \ha {{1\ov 2}}

\def \lr { \lref}
\def\np {{\it Nucl. Phys. }}
\def \pl {{\it  Phys. Lett. }}
\def \mpl {{\it Mod. Phys. Lett. }}
\def \prl {{\it  Phys. Rev. Lett. }}
\def \pr  {{\it Phys. Rev. }}

\def \cqg {{\it Class. Quant. Grav. }}

%--------+---------+---------+---------+---------+---------+---------+

%Blank macros:
\def\comment#1{}
\def\fixit#1{}

%For controlling the size of fractions:
\def\tf#1#2{{\textstyle{#1 \over #2}}}

%More math operators: (add as needed)

\def\diag{\mathop{\rm diag}\nolimits}
\def\tr{\mathop{\rm Tr}\nolimits}

%To produce a box for a Dalembertian (adapted from p. 320 of TeXbook):
\def\sqr#1#2{{\vcenter{\vbox{\hrule height.#2pt
         \hbox{\vrule width.#2pt height#1pt \kern#1pt
            \vrule width.#2pt}
         \hrule height.#2pt}}}}

%Macros to facilitate use of halign for complicated equations:
\def\TL{\hfil$\displaystyle{##}$}
\def\TR{$\displaystyle{{}##}$\hfil}

\def\TT{\hbox{##}}
%Example:
%  \eqn\One{\vcenter{\openup1\jot
%    \halign{\strut\span\TL & \span\TR & \span\TT & \span\TL & \span\TR\cr
%     x^2 &> 1 & \quad when $x$ satisfies\ \ & x &> 1 \cr
%   }}}

%List references as a continuation of the present page:

%Now, some miscellany:
%Added for partial waves paper:
%\def\l{\ell}

\def\+{^\dagger}

%Added for exact coefficients notes:

%Added for partial waves into 3-brane notes:
\def\M{\cal M}

%Added for absorption by extremal 3-branes paper.  Adapted from the 
%TeXbook, p. 359.
\def\overleftrightarrow#1{\vbox{Ialign{##\crcr
     \leftrightarrow\crcr\noalign{\kern-0pt\nointerlineskip}
     $\hfil\displaystyle{#1}\hfil$\crcr}}}

%--------+---------+---------+---------+---------+---------+---------+
%Title page

\Title{
 \vbox{\baselineskip10pt
  \hbox{PUPT-1685}
\hbox{Imperial/TP/96-97/33}
  \hbox{hep-th/9703040}
 }
}
{
 \vbox{
  \centerline{ String Theory and}
  \vskip 0.1 truein
  \centerline{Classical Absorption by Threebranes}
 }
}
\vskip -30 true pt

\centerline{
 Steven S.~Gubser,\footnote{$^1$}{e-mail: ssgubser@viper.princeton.edu} 
 Igor R.~Klebanov,\footnote{$^2$}{e-mail: klebanov@viper.princeton.edu}
 }
\centerline{\it Joseph Henry Laboratories, 
Princeton University, Princeton, NJ  08544}

\centerline{ and}

\centerline{
 Arkady A.~Tseytlin\footnote{$^3$}
  {e-mail: tseytlin@ic.ac.uk}\footnote{$^{\dagger}$}
  {Also at Lebedev Physics Institute, Moscow.} }
\centerline{{\it Blackett
             Laboratory,  Imperial College,  London SW7 2BZ, U.K.}}
\medskip
\centerline {\bf Abstract}
\medskip
\baselineskip 11pt

Low energy absorption cross-sections for various particles falling
into extreme non-dilatonic branes are calculated using string theory
and world-volume field theory methods.  The results are compared with
classical absorption by the corresponding gravitational backgrounds.
For the self-dual threebrane, earlier work by one of us demonstrated
precise agreement of the absorption cross-sections for the dilaton,
and here we extend the result to Ramond-Ramond scalars and to
gravitons polarized parallel to the brane.  In string theory, the only
absorption channel available to dilatons and Ramond-Ramond scalars at
leading order is conversion into a pair of gauge bosons on the
threebrane. For gravitons polarized parallel to the brane, scalars,
fermions and gauge bosons all make leading order contributions to the
cross-section, which remarkably add up to the value predicted by
classical gravity.  For the twobrane and fivebrane of M-theory,
numerical coefficients fail to agree, signalling our lack of a precise
understanding of the world-volume theory for large numbers of
coincident branes.  In many cases, we note a remarkable isotropy in
the final state particle flux within the brane.  We also consider the
generalization to higher partial waves of minimally coupled scalars.
We demonstrate agreement for the threebrane at $\ell = 1$ and indicate
that further work is necessary to understand $\ell > 1$.

\Date{March 1997}

\noblackbox
\baselineskip 14pt plus 1pt minus 1pt

%--------+---------+---------+---------+---------+---------+---------+
%Bibliography

\lref\ATT{A.A.~Tseytlin, {\it Mod.~Phys.~Lett.}~A11 (1996) 689,
hep-th/9601177.}

\lref\Ark{A. Tseytlin, \np Bhep-th/9602064.}

\lref\CGKT{
C.G.~Callan, Jr., S.S.~Gubser, I.R.~Klebanov and A.A.~Tseytlin,
hep-th/9610172; 
I.R. Klebanov and M. Krasnitz, hep-th/9612051.}

\lref\CTT{M. Cveti\v c and  A.A.  Tseytlin, 
\pl B366 (1996) 95, hep-th/9510097; \pr D53 (1996) 5619, 
hep-th/9512031.}

\lref\CT{M.~Cveti\v c and A.A.~Tseytlin, {\it Nucl.~Phys.}~B478 (1996)
181, hep-th/9606033.}

\lref\CYY{M.~Cveti\v c and D.~Youm, {\it Nucl.~Phys.}~B476 (1996) 118,
hep-th/9603100.}

\lref\CY{M.~Cveti\v c and D.~Youm, {\it Phys.~Rev.}~D53 (1996) 584,
hep-th/9507090.}
% Contribution to `Strings 95', hep-th/9508058.}

\lref\Duff{M.J. Duff, H. L\"u and  C.N. Pope
\pl B382 (1996) 73, hep-th/9604052.}

%\lref\EW{E. Witten, \np B460 (1996) 541.}

\lref\GKP{S.S.~Gubser, I.R.~Klebanov and A.W.~Peet, \pr D54 (1996) 3915,
hep-th/9602135.}

\lref\GKtwo{S.S.~Gubser and I.R.~Klebanov, \prl 77
(1996) 4491, hep-th/9609076.}

\lref\GK{S.S.~Gubser and I.R.~Klebanov, 
\np B482 (1996) 173, hep-th/9608108.}

\lref\GR{I.S.~Gradshteyn and I.M.~Ryzhik, {\it Table of Integrals,
Series, and Products}, Fifth Edition, A.~Jeffrey, ed. (Academic Press:
San Diego, 1994).}

\lref\HK{A. Hashimoto and I.R. Klebanov, 
\pl B381 (1996) 437, hep-th/9604065.}

\lref\HM{G.~Horowitz and A.~Strominger, {\it Phys.~Rev.~Lett.}~77
(1996) 2368, hep-th/9602051.}

\lref\HP{G. Horowitz and J. Polchinski, hep-th/9612146.}

\lref\IK{ I.R. Klebanov, hep-th/9702076.}

\lref\JP{J. Polchinski, \prl 75 (1995) 4724, hep-th/9510017.}

\lref\KT{I.R. Klebanov and A.A. Tseytlin, \np B475 (1996) 179, 
hep-th/9604166.}

\lref\LS{L. Susskind, hep-th/9309145.}

\lref\LWW{F.~Larsen and F.~Wilczek, {\it Phys.~Lett.}~B375 (1996) 37,
hep-th/9511064; hep-th/9609084.}

\lref\Unruh{W.G.~Unruh, {\it Phys.~Rev.}~D14 (1976) 3251.}

\lref\Wald{R.M.~Wald, {\it General Relativity} (Chicago: The
University of Chicago Press, 1984).}

\lref\Wass{A.~Wasserman, Part~III lectures given at Cambridge
University, Michaelmas 1995.}

\lref\age{D.~Page, {\it Phys.~Rev.}~D13 (1976) 198; {\it
Phys.~Rev.}~D14 (1976) 3260.}

\lref\aki{A.~Hashimoto, 
hep-th/9608127.}

\lref\at{A.A.~Tseytlin, {\it Nucl.~Phys.}~B475 (1996) 149,
hep-th/9604035.}

\lref\bd{P.C.W.~Birrell and N.D.~Davies, {\it Quantum Fields in Curved
Space} (Cambridge, UK: Cambridge University Press, 1982).}

\lref\bho{E.~Bergshoeff, C.~Hull and T.~Ort\'in, {\it
Nucl.~Phys.}~B451 (1995) 547, hep-th/9504081.}

\lref\bl{V. Balasubramanian and F. Larsen, hep-th/9604189.}

\lref\brek{J.C.~Breckenridge, R.C.~Myers, A.W.~Peet and C.~Vafa,
hep-th/9602065; J.C.~Breckenridge, D.A.~Lowe, R.C.~Myers,
A.W.~Peet, A.~Strominger and C.~Vafa, {\it Phys.~Lett.}~B381 (1996)
423, hep-th/9603078.}

\lref\ceder{M.~Cederwall, A.~von~Gussich, B.E.W.~Nilsson, and
A.~Westerberg, 
hep-th/9610148; M.~Cederwall, A.~von~Gussich, B.E.W.~Nilsson,
P.~Sundell, and A.~Westerberg, hep-th/9611159.}

\lref\cm{C.G.~Callan and J.M.~Maldacena, {\it Nucl.~Phys.}~B472 (1996)
591, hep-th/9602043.}

\lref\dgm{S.~Das, G.~Gibbons and S.~Mathur, hep-th/9609052.}

\lref\dmII{S. Das and S.D. Mathur,  hep-th/9607149.}

\lref\dmI{S.R.~Das and S.D.~Mathur, {\it Phys.~Lett.}~B375 (1996) 103,
hep-th/9601152.}

\lref\dmw{A.~Dhar, G.~Mandal and S.~R.~Wadia, {Phys.~Lett.}~B388
(1996) 51, hep-th/9605234.}

\lref\dm{S.R.~Das and S.D.~Mathur, {\it Nucl.~Phys.}~B478 (1996) 561,
hep-th/9606185; hep-th/9607149.}

\lref\dowk{F. Dowker, D. Kastor and J. Traschen, hep-th/9702109.}

\lref\ds{M.J. Duff and K.S. Stelle,  \pl B253
(1991) 113.}

\lref\edb{E.~Witten, {\it Nucl.~Phys.}~B460 (1996) 335.}

\lref\fkk{S.~Ferrara and R.~Kallosh, {\it Phys.~Rev.}~D54 (1996) 1514,
hep-th/9602136; {\it Phys.~Rev.}~D54 (1996) 1525, hep-th/9603090;
S.~Ferrara, R.~Kallosh, A.~Strominger, {\it Phys.~Rev.}~D{52} (1995)
5412, hep-th/9508072.}

\lref\gibb{G.~Gibbons, {\it Nucl.~Phys.}~{B207} (1982) 337;
P.~Breitenlohner, D.~Maison and G.~Gibbons, {\it
Commun.~Math.~Phys.}~120 (1988) 295.}

\lref\gkk{G.~Gibbons, R.~Kallosh and B.~Kol, hep-th/9607108.}

\lref\gkt{J.P.~Gauntlett, D.~Kastor and J.~Traschen, {\it
Nucl.~Phys.}~B478 (1996) 544, hep-th/9604179.}

\lref\gm{R. Garousi and R. Myers, hep-th/9603194.} 

\lref\gunp{S.S.~Gubser, November~1996, unpublished notes.}

\lref\guv{R. G\" uven, \pl B276 (1992) 49.}

\lref\hawk{S. Hawking and M. Taylor-Robinson, hep-th/9702045.}

\lref\hmf{{\it Handbook of Mathematical Functions}, M.~Abramowitz and
I.A.~Stegun, eds. (US Government Printing Office, Washington, DC,
1964) 538ff.}

\lref\hms{G.~Horowitz, J.~Maldacena and A.~Strominger, {\it
Phys.~Lett.}~B383 (1996) 151, hep-th/9603109.}

\lref\hrs{E.~Halyo, B.~Kol, A.~Rajaraman and L.~Susskind,
hep-th/9609075; E.~Halyo, hep-th/9610068.}

\lref\hs{G.~Horowitz  and A.~Strominger, \np B360 (1991) 197.}

\lref\jpTASI{J.~Polchinski, hep-th/9611050.}

\lref\juanI{J.~Maldacena, hep-th/9611125.}

\lref\juan{J.~Maldacena, {\it Nucl.~Phys.}~B477 (1996) 168,
hep-th/9605016.}

\lref\kaaa{R.~Kallosh, A.~Linde, T.~Ort\'in, A.~Peet and A.~Van
Proeyen, {\it Phys.~Rev.}~D{46} (1992) 5278.}

\lref\khuri{R.~Khuri, {\it Nucl.~Phys.}~B376 (1992) 350.}

\lref\km{I.R.~Klebanov and S.D.~Mathur, hep-th/9701187.}

\lref\kr{B.~Kol and A.~Rajaraman, hep-th/9608126.}

\lref\ktI{I.R.~Klebanov and A.A.~Tseytlin, 
{\it Nucl.~Phys.}~B479 (1996) 319, hep-th/9607107.}

\lref\kt{I.R. Klebanov and A.A. Tseytlin, \np B475 (1996) 165, 
hep-th/9604089.}

\lref\lu{J.X.~Lu, {\it Phys.~Lett.}~B313 (1993) 29, hep-th/9304159.}

\lref\maha{J.~Maharana and J.H.~Schwarz, {\it Nucl.~Phys.}~B390 (1993)
3, hep-th/9207016.}

\lref\mastI{J.M.~Maldacena and A.~Strominger, hep-th/9702015.}

\lref\mast{J.M.~Maldacena and A.~Strominger, hep-th/9609026.}

\lref\mst{J.M.~Maldacena and A.~Strominger, {\it Phys.~Rev.~Lett.}~77
(1996) 428, hep-th/9603060.}

\lref\ms{J.M.~Maldacena and L.~Susskind, 
{\it Nucl.~Phys.}~B475 (1996) 679, hep-th/9604042.}

\lref\myers{C. Johnson, R. Khuri and R. Myers, 
Phys. Lett. B378 (1996) 78, hep-th/9603061.} 

\lref\pertuu{G.~Gilbert, hep-th/9108012; C.F.E.~Holzhey and
F.~Wilczek, {\it Nucl.~Phys.}~B380 (1992) 447, hep-th/9202014;
R.~Gregory and R.~Laflamme, {\it Phys.~Rev.}~D51 (1995) 305,
hep-th/9410050; {\it Nucl.~Phys.}~B428 (1994) 399, hep-th/9404071.}

\lref\pktIII{P.K.~Townsend, Part~III lectures given at Cambridge
University, Lent 1995.}

\lref\pope{H. L\"u, S. Mukherji, C. Pope and J. Rahmfeld, hep-th/9604127.}

\lref\schw{J.H.~Schwarz, {\it Nucl.~Phys.}~B226 (1983) 269.}

\lref\sv{A.~Strominger and C.~Vafa, {\it Phys.~Lett.}~B379 (1996) 99,
hep-th/9601029.}

\lref\teukI{S.A.~Teukolsky, {\it Astrophys.~J.}~{\bf 185} (1973) 635.}

\lref\us{I.R.~Klebanov and L.~Thorlacius,
\pl B371 (1996) 51, hep-th/9510200;
S.S.~Gubser, A.~Hashimoto, I.R.~Klebanov and J.M.~Maldacena,
\np B472 (1996) 231, hep-th/9601057.}

%%%%%%%%%%%%%%%%%%%%%%%%%%%%%%%%%%%%%%%%%%%%%%%%%
\lref\Duff{M.J. Duff, H. L\"u and  C.N. Pope,
\pl B382 (1996) 73, hep-th/9604052. }
\lr \schw{
J.H. Schwarz, \np B226 (1983) 269.}
\lr \hulb{E. Bergshoeff, C.M.  Hull  and T. Ort\'in, \np  B451 (1995) 547, hep-th/9504081.}
\lr \bbb{E. Bergshoeff,  H.J. Boonstra  and T. Ort\'in, 
\pr D53 (1996) 7206,   hep-th/9508091.}

\lr\ght{G.W. Gibbons, G.T. Horowitz and P.K. Townsend, \cqg 12 (1995) 297,
hep-th/9410073;
M.J. Duff, G.W. Gibbons and P.K. Townsend, \pl B332 (1994) 321.}

\lr\hst {G.T. Horowitz and A. Strominger, hep-th/9602051.}

\lr\cjs {E. Cremmer, B. Julia and J. Scherk, \pl B76 (1978) 409.}

\lr\dlu{M.J. Duff and J.X. Lu, \pl B273 (1991) 409. }
\lr\dbi{R.G. Leigh, \mpl  A4 (1989) 2767; 
M. Li, \np B460 (1996) 351, hep-th/9510135;
M.  Douglas, hep-th/9512077. }
\lr\tse{ A.A. Tseytlin, \np B469 (1996) 51, hep-th/9602064.   } 
\lr\jhs{M. Aganagic, C. Popescu and J.H. Schwarz, hep-th/9612080;
E. Bergshoeff and P.K. Townsend, hep-th/9611173.}

\lr\bst {E. Bergshoeff, E. Sezgin and P.K. Townsend, \pl B189 (1988) 75.  }
\lr \ffff{E. Bergshoeff, M. de Roo and T. Ort\'in, hep-th/9606118;
M. Aganagic, J. Park, C. Popescu and J.H. Schwarz, hep-th/9701166;
I. Bandos, K. Lechner, A. Nurmagambetov, P. Pasti, D. Sorokin and M. Tonin, 
hep-th/9701149.} 
\lr\modes{G.W. Gibbons and P.K. Townsend, \prl 71 (1993) 3754;
D.M. Kaplan and J. Michelson, \pr D53 (1996) 3474, hep-th/9510053.}

\lr\alvw{L. Alvarez-Gaum\'e and E. Witten, \np B234 (1983) 269. }

\lref\KT{I.R. Klebanov and A.A. Tseytlin, \np B475 (1996) 179, 
hep-th/9604166.}

\lref\bl{V. Balasubramanian and F. Larsen, \np
 B478 (1996) 199, 
hep-th/9604189.}

\lref\CY{M.~Cveti\v c and D.~Youm, {\it Phys.~Rev.}~D53 (1996) 584,
hep-th/9507090.}

\lr\mepr {A.A. Tseytlin,  hep-th/9609212;  hep-th/9702163.}

\lr\dpl {M. Douglas, J. Polchinski and A. Strominger, hep-th/9703031.}

\lref\ATT{A.A.~Tseytlin, {\it Mod.~Phys.~Lett.}~A11 (1996) 689,
hep-th/9601177.}

\lref\mdlate{M.R.~Douglas, hep-th/9604198.}

%--------+---------+---------+---------+---------+---------+---------+
%Body

%--------+---------+---------+---------+---------+---------+---------+
\newsec{Introduction}

Dirichlet branes provide an elegant embedding of 
Ramond-Ramond charged objects into string theory \JP.
The D-brane
description of the dynamics of these solitons may be compared 
with corresponding results in the semi-classical low-energy
supergravity. In particular, a counting of degeneracies for
certain intersecting D-branes reproduces, in the limit of large
charges, the Bekenstein-Hawking entropy of the corresponding
geometries \refs{\sv,\cm}.
Furthermore, calculations of emission and absorption rates for
scalar particles agree
with a simple `effective string' model for the dynamics of the
intersection \refs{\cm,\dmw,\dm,\dmII,\GK,\mast,\GKtwo,\CGKT}.
There are some signs, however, that the simplest model is
not capable of
incorporating all the complexities of black holes physics 
\refs{\km,\hawk,\dowk}. This is probably due to the fact that
the dynamics of intersecting D-branes, which is not yet
fully understood,
cannot be captured by one simple model.

Another line of development, which has proceeded in parallel to
the investigations of intersecting D-branes, concerns the
simpler configurations which involve parallel D-branes only. 
Their string theoretic description is well understood in terms of
supersymmetric $U(N)$ gauge theory on the world-volume \edb.
Pertrubative string calculations of scattering 
\refs{\us,\gm} and absorption \HK\ 
are fairly straightforward for the parallel D-branes, 
and their low-energy dynamics is summarized in the
DBI action.

To leading order in the string coupling, $N$ coincident
Dp-branes are described by $O(N^2)$ free fields in $p+1$ dimensions.
This result may be compared with the Bekenstein-Hawking entropy
of the near-extremal $p$-brane solutions in supergravity.
In \refs{\GKP,\kt} it was found the the scaling of the 
Bekenstein-Hawking entropy with the temperature agrees
with that for a massless gas in $p$ dimensions only for the
`non-dilatonic $p$-branes'. The only representative of this class
which is described by parallel D-branes is the self-dual threebrane.
Other representatives include the dyonic string in $D=6$, and
the twobranes and fivebranes of M-theory.
In \HP\ a way of reconciling the differing scalings 
for the dilatonic branes \Duff\ was proposed.\foot{
Other ideas on how to find
agreement for the dilatonic branes 
were put forward in \pope.}
Neveretheless, in \IK\ it was shown that the non-dilatonic
branes (and especially the self-dual threebrane) have a number of
special properties that allow for more detailed comparisons between
semi-classical gravity and the microscopic theory.
For example, the string theoretic calculation of the absorption 
cross-section by threebranes
for low-energy dilatons was found to agree {\it exactly}
with the classical calculation in the background of the extremal
classical geometry \IK.

What are the features that make the threebrane so special? 
In classical supergravity, the extremal threebrane is the only
RR-charged solution that is perfectly non-singular \refs{\dlu,\ght}.
For $N$ threebranes, the curvature of the classical solution is bounded
by a quantity of order
$$ {1\over \sqrt{N\kappa_{10}}} \sim {1\over \alpha'\sqrt{N g_{\rm str}}}
\ .$$ 
Thus, to suppress the string scale corrections to the classical metric,
we need to take the limit $N g_{\rm str}\rightarrow\infty $.
This fact seems to lead to a strongly coupled theory on the world-volume
and raises questions about the applicability of string perturbation
theory to macroscopic threebranes. However, in \IK\ it was shown
that the dimensionless expansion parameter that enters the
string theoretic calculation of the
absorption cross-section is actually
\eqn\param{ N\kappa_{10} \omega^4 \sim N g_{\rm str} \alpha'^2 \omega^4
\ ,}
where $\omega$ is the incident energy. Thus, we may consider
a `double scaling limit,'
\eqn\dsl{ N g_{\rm str}\rightarrow\infty\ ,
\qquad \omega^2 \alpha'\rightarrow 0\ ,
}
where the expansion parameter \param\ is kept small.
Moreover, the classical absorption cross-section is naturally
expanded in powers of $\omega^4\times {\rm (curvature)}^{-2}$, which
is the same expansion parameter \param\ as the one governing the
string theoretic description of the threebranes. The two expansions
of the cross-section thus may indeed be compared, and 
the leading term agrees exactly \IK. This 
provides strong evidence in favor of
absorption by extremal threebranes being a unitary process.
While in the classical calculation the information carried by the
dilaton seems to disappear down the infinite throat of the classical
solution, the stringy approach indicates that the information is not
lost: it is stored in the quantum state of the back-to-back massless
gauge bosons on the world-volume which are
produced by the dilaton. Subsequent decay of the threebrane
back to the ground state proceeds via annihilation of the gauge
bosons into an outgoing massless state, and there seems to be no
space for information loss.

The scenario mentioned above certainly deserves a more careful
scrutiny. In this paper we carry out further comparisons
between string theory and classical gravity of the threebrane.
In string theory there is a variety of fields that act as minimally
coupled massless scalars with respect to the transverse dimensions.
The dilaton, which was investigated in \IK, is perhaps the simplest 
one to study. In this paper we turn to other such fields: the RR
scalar and the gravitons polarized parallel to the world-volume.
We show that in classical gravity all these fields satisfy
the same equation and, therefore, are absorbed with the same rate
as the dilaton. However, their cubic couplings to the 
massless world-volume modes dictated by string theory \HK\ are
completely different. In particular, the longitudinally polarized
gravitons can turn into pairs of gauge bosons, scalars, or fermions.
Adding up these rates we find that the total cross-section has the same
universal value, in agreement with classical gravity.

Another interesting check concerns the absorption of scalars in 
partial waves higher than $\ell=0$. In \IK\ it was shown that the
scaling of the relevant cross-sections with $\omega$ and $N$ agrees
for all $\ell$. For $\ell=1$ we show that the coefficient agrees as well.
For $\ell>1$ the simplest assumption about the
effective action does not yield a coefficient
which agrees with the classical calculation. Normalization of the
effective action is a subtle matter, however, 
and we suspect that its
direct determination from string amplitudes
will yield agreement with the classical cross-sections.

While a microscopic description of RR-charged branes is
by now well known in string theory, the situation is not as simple
for the p-brane solutions of the 11-dimensional supergravity.
There is good evidence that a yet unknown M-theory underlies
their fundamental description. Comparisons with semi-classical gravity
provide consistency checks on this description. 
In \IK\ the scalings of the classical absorption cross-sections
were found to agree with the twobrane and fivebrane effective action
considerations. 
Here we calculate the absorption cross-section
of a longitudinally polarized graviton by a single fivebrane.
Using the world-volume  effective action, we find that the 
absorption cross-section is $1/4$ of the rate formally predicted
by classical gravity. In fact, one could hardly expect perfect
agreement for a single fivebrane -- the classical description is expected
to be valid only for a large number of coincident branes.
It is interesting, nevertheless, how close the two calculations come
to agreeing with each other. We also carry out a similar comparison
for a single twobrane, but find the discrepancy in
the coefficient to be far greater than in the fivebrane case.

The structure of the paper is as follows. In section 2 we 
exhibit the space-time effective actions and derive the classical
equations satisfied by various fields. In section 3 we present
the new threebrane calculations. In section 4 we carry out the
comparisons of classical 11-dimensional supergravity with 
the predictions of the M-brane world-volume actions.
We conclude in section 5.

%%%%%%%%%%%%%%%%%%%%%%%%%%%%%%%%%%%%%%%%%%%%%%%%%%%

%%%%%%%%%%%%%%%%%%%%%%%%%%%%%%%%%%%%%%%%
\newsec{Space-time effective actions and perturbations around
non-dilatonic p-brane solutions}
%%%%%%%%%%%%%%%%%%%%%%%%%%%%%%%%
%\def \l {\lambda}
%%%%%%%%%%%%%%%%%%%%%%%%%%%
\subsec{Type IIB theory perturbations near threebranes }
%%%%%%%%%%%%%%%%%%%%%%%%%%%%%%%%%%%%%%%%%%%%%%%%
The bosonic part of the field equations 
of $d=10$ type IIB supergravity, which is the 
low-energy limit of the type IIB superstring \schw,   can be 
derived from the following    action:
  \refs{\hulb,\bbb}
\eqn\efec{  S_{10} 
= {1\ov 2\k_{10}^2}  \int d^{10} x \bigg[ \sqrt{-G} \big( e^{-2\p} [ R + 4 (\del \p)^2
- { \textstyle{1\ov 12}} (\del B_2)^2 ] 
} 
$$ -  \  { \textstyle{1\ov 2}} (\del C)^2 
 - { \textstyle{1\ov 12}} (\del C_2  - C  \del B_2) ^2
- { \textstyle{1 \ov 4\cdot 5!}}  F^2_5 \big)
- { \textstyle{1\ov 2\cdot 4! \cdot (3!)^2    }}
 {\ep_{10}} C_4 \del C_2 \del B_2 + ... \bigg] \ ,  $$
where\foot{We use the following 
notation.  The signature is $(-+...+)$. 
$M,N,...$ label the  coordinate 
indices of $d=10$ or $d=11$ theory.
 The indices $\a,\b,...$ ($a,b,..$)
  will label the  space-time  (spatial) coordinates parallel
to the p-brane  world-volume, i.e. $\a =(0,a), \ a=1,...,p$;  
  the indices $i,j,...,$   
will label the coordinates transverse
to  the p-brane, $i=p+1,...,9 (10)$. We  shall  also use
$\m,\n,...$ for the indices of the dimensionally reduced
theory obtained by compactifying the internal directions
parallel to the  p-brane, 
i.e. $\m= (0,i)$. $\ep_d$ will  stand for the totally antisymmetric
 symbol (density). Contractions over repeated 
lower-case indices are always performed with the flat  metric, 
and $A_{[ab]} \equiv  \ha (A_{ab} - A_{ba})$.}
$$(\del B_2)_{MNK} \equiv 3 \del_{[M} B_{NK]}\ , \ \ \ 
\  \ (\del C_4)_{MNKLP} \equiv 5 \del_{[M} C_{NKLP]} \ , $$ 
$$ F_5= \del C_4 + {5} (B_2 \del C_2 - C_2 \del B_2) \  . $$
Following \bbb\ we assume   that the self-duality constraint 
 \ $F_5= \tilde{F}_5$
may be  added  at the level of the equations of motion.
This action is a useful tool for deriving the dimensionally reduced 
forms of type IIB supergravity action  \bbb\ and also 
for discussing perturbations near the solitonic p-brane 
solutions given below.
Written in the Einstein frame ($g_{MN} = e^{-\p/2} G_{MN}$)
it takes the following $SL(2,R)$ covariant form: 
\eqn\efc{  S_{ 10} 
= {1\ov 2\k_{10}^2}  \int d^{10} x \bigg( \sqrt{-g_{10}} \big[ \ R 
 - { \textstyle{1\ov 2}} (\del \p)^2
- { \textstyle{1\ov 12}} e^{-\p}   (\del B_2)^2 
} 
$$ -  \  { \textstyle{1\ov 2}}  e^{2 \p} (\del C)^2 
 - { \textstyle{1\ov 12}} e^{ \p}  (\del C_2  - C  \del B_2) ^2
- { \textstyle{1 \ov 4\cdot 5!}}  F^2_5\ \big]
- { \textstyle{1\ov 2\cdot 4! \cdot (3!)^2    }}
 {\ep_{10}} C_4 \del C_2 \del B_2 + ... \bigg) \ . $$
The extremal 3-brane of type IIB theory is represented
by the background \refs{\hs,\dlu}
\eqn\tree{
ds^2_{10 } = H^{-1/2}(-dt^2 + dx_a dx_a) + H^{1/2} dx_i dx_i  \ , }
\eqn\tre{ 
(\del C_4)_{abcdi} = \ep_{abcd} \del_i H\inv \ , \ \ \ \ 
(\del C_4)_{ijklm} = \ep_{ijklmn}\del_n H \ , \ \ \ \  
H= 1 + {R^4\ov r^4} \ , \ \ r^2=x_ix_i\ , }
with all other  components  and  fields  vanishing (i.e. $\p=C=B_2=C_2=0$, \ 
$F_5= \del C_4$).
The scalar curvature $\sim (\del C_4)^2  \sim R^{-2}$ near $r=0$,
In fact, this metric may be extended to a geodesically complete
 non-singular geometry \ght. This is the only RR-charged
p-brane for which this is possible.

Let us consider small perturbations near this background
which depend only on $x_\m$, i.e.  on the 
time $t$  and the transverse coordinates $x_i$.
Our aim is to identify the modes which have simple 
Klein-Gordon type equations.  A  guiding principle
is to look  for  (components of) the fields
that have trivial background values. 

The obvious examples are the  
dilaton $\p$ 
and  the Ramond-Ramond scalar $C$ perturbations which  are  decoupled
from the 
$C_4$-background and thus  have the action ($\vp=(\p,C)$) 
\eqn\scal{ S_{scal.} =- {1\ov 4\k_{10}^2} 
 \int d^{10} x  \sqrt{-g_{10}}\  g^{\m\n} \del_\m \vp \del_\n \vp 
= - {1\ov 4\k_{10}^2} 
 \int d^{10} x  \big[
   \del_i \vp \del_i \vp  -  H(r)  \del_0 \vp \del_0 \vp  \big]  , }
where we have used the fact that for the 3-brane background 
the Einstein and the string frame
metrics are identical, $G_{MN}=g_{MN}$,
which implies $ g_{\m\n} =
\diag (-H^{-1/2}, H^{1/2}\delta_{ij} )$, \ $\sqrt{-g_{10}}= H^{1/2}$.

Perturbations of the metric are, in general, mixed with
perturbations of components of  $C_4$.  An
important  exception is 
the traceless part of the longitudinal (polarized along the
3-brane) graviton  perturbations 
$h_{ab}$. These perturbations give rise to scalars upon
dimensional reduction to $d=7$. 
Let us  split the metric in the  `7+3'  fashion
$$ ds^2_{10E}= g_{\m\n} dx^\m dx^\n + g_{ab} dx^a dx^b \ ,$$
and assume that all  the fields depend only on $x^\m$
(this is equivalent to  reduction to 7 dimensions).
Then the relevant part of \efc\ becomes 
\eqn\efic{  S_{10} 
= {1\ov 2\k_{10}^2}  \int d^{10} x  \sqrt{-g_7} \sqrt {g_3}  
\big( R_7 
 - { \textstyle{1\ov 4}} g^{ab} g^{cd} 
 \del_\m  g_{ac}\del^\m  g_{bd}
- { \textstyle{1 \ov 48}}  F_{\m \n abc} F^{\m \n}_{ \ \  \  def} 
g^{ad} g^{be}  g^{cf}  + ... \big) , }
where $g_7 \equiv \det g_{\m\n}, 
\ g_3\equiv \det {g_{ab}}$ 
and we have written down explicitly the only 
$F^2_5$ term that could potentially couple
$h_{ab}$ to the gauge field strength
background.  Introducing 
the `normalized' metric $\g_{ab} =  g_3^{-1/3} g_{ab}$
which has unit determinant, we find
\eqn\ecf{  S_{scal.grav.} 
= {1\ov 2\k_{10}^2}  \int d^{10} x  \sqrt{-g_7} \sqrt{ g_3 } 
\big( R_7 
 - { \textstyle{1\ov 4}} \g^{ab} \g^{cd} 
 \del_\m  \g_{ac}\del^\m  \g_{bd}   }
$$ 
- \  { \textstyle{1\ov 12}} g_3^{-2}  \del_\m g_3 \del^\m g_3 
- { \textstyle{1 \ov 8}}  g_3^{-1}
 F_{\m \n} F^{\m \n} + ... \big) \ , $$
where $F_{\m\n} = F_{\m\n 123}$ is the $d=7$ vector field strength
which, according to \tre, describes   an  electrically
charged extremal black hole. Thus, only
the determinant of $g_{ab}$, which is related to
the 7-dimensional dilaton, 
couples to the gauge field 
background. 
This  is consistent with  
the fact that $\g_{ab}$ has a trivial background value,
$\g_{ab} =\delta_{ab}$,  while the value of $\det g_{ab}$ in \tree\ is 
\  $g_3= H^{-3/2}$. 

The fluctuations  $h_{ab}  = \g_{ab}- \delta_{ab}$ 
(which are traceless, $h_{aa}= 0$, to keep $\det \g_{ab}=1$) 
 thus have the 
same  quadratic part of the action as the scalar fields in \scal\
\eqn\ligr{ S_{scal.grav.} =- {1\ov 8\k_{10}^2} 
 \int d^{10} x  \sqrt{-g_{10}}\  \del_\m h_{ab} \del^\m h_{ab} } $$
=  - {1\ov 8 \k_{10}^2} 
 \int d^{10} x  \big[
   \del_i h_{ab} \del_i h_{ab}  -  H(r)  \del_0 h_{ab} \del_0 h_{ab}  \big]  \ . $$
Similar conclusions hold for other p-brane solutions because
the `normalized' internal space
metric  $\g_{ab}$ has a flat background value.

The problem of discussing perturbations that have higher spin
from the point of view of the $d=7$ black hole is more difficult
because of mixing among different fields.
We include some observations on this problem in Appendix.

%%%%%%%%%%%%%%%%%%%%%%%%%%%%%%%%%%%%%%%%%%%%%%%%%%%%%
\subsec{$d=11$ supergravity perturbations near twobranes and fivebranes}
%%%%%%%%%%%%%%%%%%%%%%%%%%%%%%%%%%%%%%%%%%%%%%%%
The  discussion of perturbations around the twobrane and fivebrane solutions
 \refs{\ds,\guv}
of the $d=11$ supergravity  is very similar to the analysis in the
preceding subsection. 
The starting point is the bosonic  part of the  
$d=11$ supergravity action \cjs\ 
\eqn\eleva{  S_{11} 
= {1\ov 2\k_{11}^2}  \int d^{11} x \bigg( \sqrt{-g_{11}} \big[ R 
 - { \textstyle{1\ov 2\cdot 4!}} (\del C_3)^2 \big] 
+  { \textstyle{1\ov (12)^4 }}  {\ep_{11}} C_3 \del C_3 \del C_3 + ... \bigg) \ , }
where 
$
(\del C_3)_{MNKL} = 4 \del_{[M} C_{NKL]}$. 
The twobrane and fivebrane
backgrounds  are respectively ($a=1,...,p, \ i=p+1,...,10$, \ $p=2,5$)
\eqn\twe{
ds^2_{11}  = H^{-2/3}(-dt^2 + dx_a dx_a) + H^{1/3} dx_i dx_i  \ , }
$$ 
(\del C_3)_{0abn} = \ep_{ab}\del_n H\inv  \ , \ \ \ \  
H= 1 + {R^6\ov r^6} \ ,$$
\eqn\five{
ds^2_{11}  = H^{-1/3}(-dt^2 + dx_a dx_a) + H^{2/3} dx_i dx_i  \ , }
$$
(\del C_3)_{ijkl} = \ep_{ijkln}\del_n H \ , \ \ \ \  
H= 1 + {R^3\ov r^3} \ .$$
Both for the twobrane and for the fivebrane, the
scalar curvature is $\sim (\del C_3)^2  \sim R^{-2}$ near $r=0$.
There is a subtle difference, however, in that the fivebrane
metric may be extended to a geodesically complete non-singular
geometry, while the twobrane metric  cannot \ght.

We consider small perturbations near these backgrounds
which depend only on $x_\m=(t,x_i)$.
There are no scalars like $\p$ or $C$ in the $d=11$ theory, but 
it is easy to check that,  as  in \efic,\ecf,  the
`normalized' internal part of the metric 
 $\g_{ab} = g_p^{-1/p} g_{ab}, \   g_p\equiv \det g_{ab}$
($p=2,5$),   which has a flat background value, 
 is decoupled from the $(\del C_3)^2$
term. As a result, the gravitons polarized parallel to the brane,
$h_{ab} = \g_{ab} - \delta_{ab}, \ h_{aa}=0,$
have the minimal action  as in \ligr, 
\eqn\lgr{ S_{scal.grav.} =- {1\ov 8 \k_{11}^2} 
 \int d^{11} x  \big[
  \del_i h_{ab} \del_i h_{ab}  -  H(r)  \del_0 h_{ab} 
\del_0 h_{ab}  \big] \ . }
We discuss some other `mixed' perturbations in  Appendix.

%%%%%%%%%%%%%%%%%%%%%%%%%%%%%%%%%%%%%%%%%%%%%%%%
\newsec{The Self-Dual Threebrane}
%%%%%%%%%%%%%%%%%%%%%%%%%%%%%%%%%%%%%%

%%%%%%%%%%%%%%%%%%%%%%%%%%%%%%%%%%%%%%%%%%%%%%%%%%%
\subsec{Classical absorption}

This section focuses on minimally coupled scalars, such as
those that satisfy the Klein-Gordon equation following from \scal, \ligr\
with $H(r)$ given in  \tre. 
Because the classical analysis of absorption 
for all partial waves of such a
scalar was sketched in \IK\ and is similar to the matching
calculations that have appeared in many other places in the literature
\refs{\dmw,\dm,\GK,\mast,\GKtwo,\CGKT,\Unruh,\age,\dgm,\kr},
details will not be presented here.
In order to employ analytical rather than numerical techniques, it is
necessary as usual to take the Compton wavelength much larger than
the typical radii of the black hole.  Fortunately, it is precisely in
this region where agreement with string theoretic models is
expected~\juanI.

It is useful to know the Optical Theorem in arbitrary spacetime
dimension $d$, which can be derived from properties of the partial
wave expansion in arbitrary dimension~\gunp.  The result we need
is that the absorption cross-section $\sigma^\ell_{\rm abs}$ of a
massless scalar with energy $\omega$ 
in the $\ell$-th partial wave is related to the
absorption probability $1 - |S_\ell|^2$ by
  \eqn\SigmaAbsGen{
   \sigma^\ell_{\rm abs} = {2^{d-3} \pi^{(d-3)/2} \over \omega^{d-2}}
    \Gamma((d-3)/2) (\ell + (d-3)/2) {\ell+d-4 \choose \ell} 
    \left( 1 - |S_\ell|^2 \right) \ .
  }
 For the threebrane one takes $d=7$ since the other $3$~dimensions are
compactified on $T^3$.  
A matching calculation outlined for all partial waves in
\IK\ gives the absorption probability
$$ 1 - |S_\ell|^2= {\pi^2 (\omega R)^{8+4\ell}\over
[(\ell+1)!]^4 (\ell+2)^2 4^{2\ell+2}}
\ .$$
Therefore, the classical
result for the cross-section to absorb the $\ell$-th partial
wave of a minimally coupled scalar 
is, to leading order in $(\omega R)^4$,
  \eqn\SigmaGR{
   \sigma_{3{\rm \ class.}}^\ell = {\pi^4 \over 24} 
    {(\ell+3) (\ell+1) \over [(\ell+1)!]^4} 
    \left( {\omega R \over 2} \right)^{4\ell} \omega^3 R^8 \ . 
  }
The  scale  parameter $R$  of the classical 3-brane solution \tre\ 
is related \refs{\GKP,\IK} to the number $N$ of coinciding microscopic 
3-branes
by the equation 
\eqn\chaq{ R^4 = {\kappa_{10}\over 2\pi^{5/2}} N \ , 
}
which follows from the quantization of the threebrane charge.

There is a number of minimally coupled scalars in the theory: the
dilaton, the RR scalar, and `off-diagonal'  gravitons polarized with both
indices parallel to the 3-brane world-volume.  The goal of the next
section will be to demonstrate that the universality of leading order
cross-section for these scalars, which is so obvious in the
semi-classical framework, also follows from the D-brane description.

%%%%%%%%%%%%%%%%%%%%%%%%%%%%%%%%%%%%%%%%%%%%%%%%%%%%%%%%%%%%%%%%%%%%
\subsec{Universality of the absorption cross-section
for minimally coupled scalars}
%%%%%%%%%%%%%%%%%%%%%%%%%%%%%%%%%%%%%%%%%%%%%%%%%%%%%%%%%%%%%%%%%%%%%%%

The threebrane is the case where we know the world-volume theory the
best: at low energies (and in flat space) 
 it is ${\cal N}=4$ supersymmetric $U(N)$ gauge theory 
where $N$ is the number of parallel threebranes 
\edb.  Thus, the massless fields on the
world-volume are the gauge field, 6 scalars, and 4 Majorana fermions,
all in the adjoint representation of $U(N)$. 
As we will see below, the universality of the cross-section is
not trivial in the world-volume description:
while for dilatons and RR scalars leading absorption proceeds by
conversion into a pair of gauge bosons only, for the
gravitons polarized along the brane
it involves a summation over conversions into
world-volume scalars, fermions, and gauge bosons.

The world-volume action, excluding all couplings to external fields, is
($I=1,...,4; \ i=4,...,9$)
  \eqn\NFourS{
   S_3 = T_3 \int d^4 x \, \tr \left[ -\tf{1}{4} F_{\alpha\beta}^2 + 
    \tf{i}{2}
%\sum_{I=1}^4
 \bar\psi^I \gamma^\alpha \partial_\alpha \psi_I - 
    \tf{1}{2}
%\sum_{i=4}^9
 (\partial_\alpha X^i)^2 + {\rm interactions} \right] \ .
  }
 As is clear from the $d=10$ origin of this theory, the R-symmetry
group is the group $SO(6)$ of spatial rotations in the uncompactified
dimensions.  Under this group, the gauge fields are neutral and the
scalars $X^i$ form a ${\bf 6}$.  When the fermions are written in
chiral components,
  $
   \psi_I = {\lambda_I \choose \bar\lambda^I}
  $,
 the fields $\lambda_I$ form a ${\bf 4}$ of $SO(6) = SU(4)$.

In order to properly normalize the amplitudes, one needs the kinetic
part of the action for the bulk fields, which is given by
\scal\ and \ligr.
%  For off-diagonal gravitons,
%dilatons, and RR scalars, this action is
%  \eqn\Sbulk{
%   S = {1 \over 2 \kappa_{10}^2} \int d^{10} x \,
%    \left( \tf{1}{4} (\partial_L h_{MN})^2 - 
%     \tf{1}{2} (\partial_M \phi)^2 - 
%     \tf{1}{2} (\partial_M C)^2 \right) \ .  }
% The first term in \Sbulk\ arises from writing 
%  $g_{MN} = \eta_{MN} + h_{MN}$.

We also need to know how the threebrane
world-volume fields couple to the bulk fields of type~IIB
supergravity.  For a single threebrane,  a $\k$-symmetric version of the
DBI action in a non-trivial type IIB background 
was constructed in \ceder\ (see also \refs{\jhs}),
which, by use of superfields, captures 
all such couplings to leading order in derivatives
of external fields. 
For the terms necessary to us, the generalization from $U(1)$ to
$U(N)$ gauge group is straightforward. One can, in principle, 
obtain  detailed  information  about the structure of the non-abelian 
action by directly computing
string amplitudes as done in \HK.  The threebrane is
well suited to this line of attack because the string theory
description is known exactly and is not
complicated by the difficulties one encounters 
in bound states of intersecting branes, such as
the D1-D5-brane system \aki.

$SO(6)$ invariance and power counting in the string coupling greatly
restrict the possible couplings between the bulk and world-volume
fields in the interaction part of the action $S_{\rm int}$.  In both
the semi-classical and world-volume computations, the dimensionless
expansion parameter is not $N g_{\rm str}$ but rather $(\omega R)^4$.
A cross-section which involves $n$ powers of this parameter arises
from a coupling of the bulk field to a local operator ${\cal O}$ of
dimension $n+4$.

The leading   bosonic terms in the action for a  single 3-brane 
in a type IIB supergravity background  are \refs{\dbi,\tse}
\eqn\teea{
 S_3=-T_3 \int d^4 x \bigg(
    \sqrt {-\det (\hat  g  +  e^{-\p/2} \F) }
    +{\textstyle  {1\ov 4!} } \ep^{\a\b\s\r} \hat C_{\a\b\s\r} 
+   \tf{1}{2}    \hat C_{\a\b} \td \F^{\a\b}
+  \four  C   \F_{\a\b}\td  \F^{\a\b} \bigg) 
 \ , }
where 
$$ \F_{\a\b}= F _{\a\b}+ \hat B_{\a\b}\ , \qquad
\td F^{\a\b} = {{\textstyle{1\over 2}}} \ep^{\a\b\s\r}F_{\s\r}\ , \qquad
\hat g_{\a\b} = g_{MN} \del_\a X^M \del_\b X^N \ , {\rm etc.}, 
$$
and the background fields are functions of $X^M$.
In the static gauge ($X^\a = x^\a$, $\a=0,\ldots, 3$) one has 
$$\hat g_{\a\b} = g_{\a\b} + 2g_{i(\a } \del_{\b )} X^i +
 g_{ij } \del_{\a} X^i\del_{\b} X^j 
\ .$$
In flat space the  fermionic terms 
may be found  by replacing  
$\hat  g_{\a\b}  +  e^{-\p/2} \F_{\a\b}  $ by \refs{\ceder,\jhs} 
$$\hat  g_{\a\b }  +  e^{-\p/2}[ \F_{\a\b}  -  
2 \bar \Psi (\Gamma_\a + \Gamma_i 
\del_\a X^i)\del_\b  \Psi  + \bar \Psi \Gamma^i \del_\a  \Psi 
\bar \Psi \Gamma_i \del_\b  \Psi]\ , $$
where $\Psi$ 
is  the $d=10$ Majorana-Weyl spinor (which can be split into 
four $d=4$ Majorana spinors $\psi_I$ in \NFourS) 
and $\Gamma_M$ are the $d=10$ Dirac matrices. 

The  leading-order 
interaction of the dilaton with world-volume fields 
implied by \teea\   was discussed 
in \IK.  The coupling of the RR scalar $C$  is  
similar, being related by $SL(2,R)$ 
duality.
The coupling of  the gravitons polarized parallel to the brane, 
$h_{\alpha\beta}=g_{\a\b}-\eta_{\a\b}$,
 to the bosonic world-volume fields can be
deduced  by expanding the action \teea.
 At  the leading order  it is  given by 
  $\tf{1}{2} h^{\alpha\beta} T^{\rm bosons}_{\alpha\beta}$
 where $T^{\rm bosons}_{\alpha\beta}$ is  the energy-momentum
tensor  of $A_\a$ and $X^i$. The 
  only possible supersymmetric extension
is for $h_{\alpha\beta}$ to couple to the complete energy-momentum
tensor corresponding to \NFourS.

Generalizing to $U(N)$, we find that
the part of $S_{\rm int}$ that is relevant to the leading-order 
absorption processes we wish to consider is\foot{We  ignore 
the  fermionic couplings like $ \phi  \bar\psi^I \gamma^\alpha
 \partial_{\a} \psi_I $ and  similar ones for $C$ and $h_{\a\b}$ 
 which are proportional to the 
fermionic equations of motion and thus  
give vanishing contribution to the S-matrix elements.}
\eqn\RelSint{
   S_{\rm int} = T_3 \int d^4 x \, \left[ \tr \left(
 \tf{1}{4}   {\phi} F_{\alpha\beta}^2  -
  \tf{1}{4}   {C} F_{\alpha\beta} \td {F}^{\alpha\beta} \right) +  
     \tf{1}{2} h_{\alpha\beta} T_{\alpha\beta} \right] \ , 
  }
 where 
  \eqn\Tab{
   T_{\alpha\beta} = \tr \big[ F_{\alpha}^{\ \gamma} F_{\beta\gamma} - 
    \tf{1}{4} \eta_{\alpha\beta} F_{\gamma\delta}^2 -
   \tf{i}{2} 
%\sum_{I=1}^4 
\bar\psi^I \gamma_{(\alpha} \partial_{\beta)} \psi_I 
  +
% \sum_{i=4}^9 \big[
\partial_\alpha X^i \partial_\beta X^i  - 
     \tf{1}{2} \eta_{\alpha\beta} (\partial_\gamma X^i)^2
% \big] 
\big] \ .
  }

Let us first consider an off-diagonal graviton
polarized along the brane, say
$h_{xy}$, which is an  example of a traceless
perturbation $h_{ab} $
 whose quadratic action is given in \ligr). {}From \RelSint\ one
 can read off the invariant amplitudes
for absorption into two scalars, two fermions, or two gauge bosons:
  \eqn\InvAmps{\vcenter{\openup1\jot
   \halign{\strut\span\TT & \span\TL & \span\TR\cr
    scalars:\ \  \ & {\cal M} &= -\sqrt{2 \kappa_{10}^2} 
     (p_{1x} p_{2y} + p_{1y} p_{2x})  \cr
    fermions:\ \ \ & {\cal M} &= -\tf{1}{2} \sqrt{2 \kappa_{10}^2}
     \bar{v}(-p_1) (\gamma_x p_{2y} + \gamma_y p_{2x}) u(p_2) \cr
    gauge bosons:\ \ \ & {\cal M} &= -\sqrt{2 \kappa_{10}^2} 
      (f^{(1)}_x{}^\beta f^{(2)}_{y\beta} + 
      f^{(1)}_y{}^\beta f^{(2)}_{x\beta})  \cr
  }}}
 where $p_1$ and $p_2$ are the momenta of the outgoing particles, and 
the field strength polarization tensors $f^{(s)}_{\alpha\beta}$ are
given by 
  \eqn\fDef{
   f^{(s)}_{\alpha\beta} = 
    i p_{s\alpha} \epsilon^{(s)}_\beta - 
    i p_{s\beta} \epsilon^{(s)}_\alpha \ .
  }
 Summation over the spins of the outgoing particles can be performed
using
\eqn\SpinSums{
  \sum_s u_{(s)}(p) \bar{u}_{(s)}(p) = p\!\!\!/ \ , \ \ \ \  \ 
  \sum_s v_{(s)}(p) \bar{v}_{(s)}(p) = -p\!\!\!/ \ , }
$$
  \sum_s \epsilon_\alpha^{(s)}  \epsilon_{\beta}^{(s)*} = 
    \eta^{\alpha\beta} 
\ .$$
 Summing as well over different species of particles available (six
different $X^i$, for example), one obtains
  \eqn\AvAmpSq{\vcenter{\openup1\jot
   \halign{\strut\span\TT & \span\TL & \span\TR\cr
    scalars:\ \ \ & \overline{|{\cal M}|}^2 &= 
     3 \kappa_{10}^2 \omega^4 n_x^2 n_y^2  \cr
    fermions:\ \ \ & \overline{|{\cal M}|}^2 &= 
     \kappa_{10}^2 \omega^4 (n_x^2 + n_y^2 - 4 n_x^2 n_y^2)  \cr
    gauge bosons:\ \ \ & \overline{|{\cal M}|}^2 &= 
     \kappa_{10}^2 \omega^4 (1 - n_x^2 - n_y^2 + n_x^2 n_y^2) \cr
  }}}
 where $\vec{n}$ is the direction of one of the outgoing particles.
In \AvAmpSq\ we have anticipated conservation of energy and momentum by
setting $\vec{p}_1 + \vec{p}_2 = 0$ and $\omega_1 + \omega_2 =
\omega$. 
 It is remarkable that the sum of these three quantities is
independent of $\vec{n}$.  Thus, if one performs
the spin sums not just over all polarizations and species of particles
of a given spin, but rather over all the states in the ${\cal N}=4$
super-multiplet, the result is isotropic:
  \eqn\TotAmpSq{
   \overline{|{\cal M}|}^2 = \kappa_{10}^2 \omega^4 \ .
  }
 The absorption cross-section is evaluated from $\overline{|{\cal
M}|}^2$ in precisely the same way that decay rates of massive
particles are calculated in conventional 4-dimensional field theories:
  \eqn\SigmaForm{
   \sigma_{3{\rm \ abs}} = {N^2 \over 2} {1 \over 2 \omega} 
    \int {d^3 p_1 \over (2 \pi)^3 2 \omega_1}
         {d^3 p_2 \over (2 \pi)^3 2 \omega_2} \,
    (2 \pi)^4 \delta^4 \big( q - {\textstyle \sum\limits_i} p_i \big) \ 
     \overline{|{\cal M}|}^2 \ .
  }
 The leading factor of $N^2$ accounts for the multiple branes; the
$1/2$ is present because the outgoing particles are identical (this is
true also of the fermions because we are working with
Majorana spinors).  The cross-section following from \TotAmpSq\ agrees
with the semi-classical $\ell = 0$ result \SigmaGR:
  \eqn\SigmaGrav{
   \sigma_{3{\rm\ abs}} = {\kappa_{10}^2  \omega^3 N^2\over 32 \pi} 
     = \sigma_{3{\rm\ class.}} \ ,
  }
where we have used the relation \chaq\ 
between $N$ and $R$.
 
Compared to the off-diagonal graviton, the calculation of the
cross-section for the RR scalar is relatively simple.  Inspection of
the leading order amplitudes for absorption of a dilaton and a RR
scalar makes it  obvious that they have the same cross-section:
  \eqn\InvAmps{\vcenter{\openup1\jot
   \halign{\strut\span\TT & \span\TL & \span\TR\cr
    dilaton:\ \ \ & {\cal M} &= \tf{1}{2} \sqrt{2 \kappa_{10}^2} 
     f^{(1)}_{\alpha\beta} f^{(2)\alpha\beta}  \cr
    RR scalar:\ \ \ & {\cal M} &= -\tf{1}{2} \sqrt{2 \kappa_{10}^2}
     f^{(1)}_{\alpha\beta} \tilde{f}^{(2)\alpha\beta} \ .  \cr
  }}}
 To show that $\overline{|{\cal M}|}^2 = \kappa_{10}^2 \omega^2$ in both
cases, it suffices to prove the relation
  \eqn\EpsRel{
   \sum_{\rm spins}
 f^{(1)}_{\alpha\beta} f^{(1)*}_{\gamma\delta}  
 f^{(2)\alpha\beta} f^{(2)\gamma\delta*} = 
    \sum_{\rm spins} 
 f^{(1)}_{\alpha\beta} f^{(1)\gamma\delta *}  
\tilde{f}^{(2)\alpha\beta} 
     \tilde{f}^{(2)\gamma\delta*} = 8 (p_1 \cdot p_2)^2 \ .
  }
 The verification is straightforward algebra.  The formula \SigmaForm\
applies as written to the RR scalar as well, so the agreement with the 
semi-classical calculation is clear.

\subsec{Higher partial waves}

A more difficult comparison is the absorption cross-section for higher
partial waves.  It is possible to argue, based on power counting and
group theory, that the interaction term suggested in \IK, 
  \eqn\LthPartial{
   S_{\rm int}^\ell = - T_3 \int d^4 x \, \ \tf{1}{4 \cdot \ell!}
    {\partial_{i_1} \cdots \partial_{i_\ell} \phi}
    \tr \left( X^{i_1} \cdots X^{i_\ell} 
      F_{\alpha\beta} F^{\alpha\beta}\right) 
  }
 is the only one that could possibly contribute at leading order to a
given partial wave. It was shown in \IK\ that this term predicts a
cross-section for the $\ell$-th partial wave whose scaling with
$\omega$ and $N$ is in agreement with classical gravity.  Here we show
that, if we use the specific normalization given in \LthPartial, we
obtain precise agreement for $\ell = 0,1$ but disagreement
for higher $\ell$.  Later on we will argue that the disagreement for
$\ell>1$ may be due simply to an incorrect normalization of the
necessary effective action terms.

Let us restrict our attention to the dilaton.  It will be obvious that
all our arguments apply equally well to the RR scalar, and perhaps
with a bit more attention to details to the off-diagonal gravitons.
The absorption cross-section \SigmaGR\ is of order $\kappa_{10}^{\ell
+ 2}$ for the $\ell$-th partial wave.  Processes which can contribute
to this absorption at leading order must be mediated by an operator
${\cal O}$ which involves $\ell + 2$ fields.  All interaction terms
which involve the dilaton must include at least two powers of
$F_{\alpha\beta}$ because of the restricted way in which the dilaton
enters the DBI action.  Since the gauge bosons are neutral under
$SO(6)$, the other $\ell$ fields must be responsible for balancing the
$SO(6)$ transformation properties of the $\ell$-th partial wave.  We
want to argue that the only way this can be done is to use $\ell$
powers of the scalar fields $X^i$.  To do this it is necessary to know
someting about the addition rules for representations of $SO(6) =
SU(4)$.  The $SU(4)$ Young tableaux for the fields in question are
  \eqn\YoungT{
   \hbox{
    \vtop{\hbox{$\lambda_I$:}}\enspace
     \vtop{\null\nointerlineskip\vskip-10pt
           \hbox{\psfig{figure=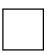}}}\qquad
    \vtop{\hbox{$\bar\lambda^I$:}}\enspace
      \vtop{\null\nointerlineskip\vskip-10pt
           \hbox{\psfig{figure=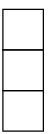}}}\qquad
    \vtop{\hbox{$X^i$:}}\enspace
     \vtop{\null\nointerlineskip\vskip-10pt
           \hbox{\psfig{figure=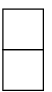}}}\qquad
    \vtop{\hbox{$\phi_\ell$:}}\enspace
     \vtop{
      \hbox{
       \vtop{\null\nointerlineskip\vskip-10pt
             \hbox{\psfig{figure=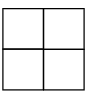}}}
       \vtop{\hbox{$\ldots$}}\enspace
       \vtop{\null\nointerlineskip\vskip-10pt
             \hbox{\psfig{figure=two.eps}}}
      }\nointerlineskip\vskip-1pt
      \hbox{$\ \displaystyle
       \underbrace{\qquad\qquad\qquad}_{\ell\rm\;columns}$}
     }
   }    
  }
 where $\phi_\ell$ stands for the $\ell$-th partial wave.  There
is a complicated general procedure known as the Littlewood-Richardson
rule for taking tensor products of representations of $SU(N)$.  It
becomes much simpler when one of the factors is a fundamental
representation ($k$ boxes in a single column) \Wass.  In that case, one
adds these $k$ boxes to the other factor's Young tableaux in all
possible ways, modulo the restriction that not more than one box can
be added to any given row.  Then one eliminates all columns
containing $N$ boxes.  Using this rule it is easy to show that with
$\ell$ fundamental representations as the factors in a tensor product,
it is impossible to obtain the tableau for $\phi_\ell$ unless all the
factors are ${\bf 6}$ representations.

A shorter argument can be made based on purely dimensional grounds.
The dimension of ${\cal O}$ must be $\ell+4$ or the cross-section
would be suppressed by extra powers of $\omega$.  Since
$F_{\alpha\beta}^2$ has dimension $4$, the other $\ell$ fields must
add $\ell$ to the total dimension of ${\cal O}$.  The only possibility
is $\ell$ scalars.

The upshot is that the only possible term in the action which can
contribute at leading order to the absorption of a dilaton in the
$\ell$-th partial wave is, up to normalization, given by
\LthPartial.
 The normalization given there is the one arising from Taylor expanding
$\phi(X)$.\foot{As we argue below, it is possible that this
local expansion is not consistent with the string amplitude
calculations. This may remove the discrepancy for
$\ell>1$.}  Actually, this action is not
quite the one we want for $\ell > 1$: in addition to mediating the
process which contributes at leading order to the absorption of the
$\ell$-th partial wave, it makes subleading contributions to
lower partial waves.  In order to isolate the $\ell$-th partial
wave, the product $X^{i_1} \cdots X^{i_\ell}$ should be replaced by an
expression which transforms irreducibly under $SO(6)$.  For $\ell = 2$
and $\ell = 3$, the appropriate replacements are
  \eqn\LTwoThree{\vcenter{\openup1\jot
   \halign{\strut\span\TT & \span\TL & \span\TR\cr
    $\ell = 2$: & X^{i_1} X^{i_2} &\to 
     X^{i_1} X^{i_2} - \tf{1}{6} \delta^{i_1 i_2} X^2  \cr
    $\ell = 3$: & \ X^{i_1} X^{i_2} X^{i_3} &\to
     X^{i_1} X^{i_2} X^{i_3} - 
      \tf{3}{8} \delta^{(i_1 i_2} X^{i_3)} X^2  \ . \cr
  }}}
 One thus subtracts an $\ell = 0$ contribution from the putative $\ell
= 2$ term in \LthPartial, and an $\ell = 1$ contribution from $\ell =
3$.  The whole Taylor series can be thus reshuffled.

Because the replacements in \LTwoThree\ project out a final state with
definite $\ell$, it is permissible to let the initial dilaton
wave-function be $e^{-i \omega (t-x^1)}$ as usual.  The derivatives
$\partial_i$ in \LthPartial\ can then be replaced by $i \omega
\delta^1_i$.  The spin-summed amplitudes squared for absorption of the
first few partial waves are 
  \eqn\FewAmps{\vcenter{\openup1\jot
   \halign{\strut\span\TT & \span\TL & \span\TR\cr
    $\ell = 0$:\ \ \ & \overline{|{\cal M}|}^2 &= 
     4 \kappa_{10}^2 (p_1 \cdot p_2)^2  \cr
    $\ell = 1$:\ \ \ & \overline{|{\cal M}|}^2 &= 
     {4 \over \sqrt\pi} \kappa_{10}^3 \omega^2 (p_1 \cdot p_2)^2  \cr
    $\ell = 2$:\ \ \ & \overline{|{\cal M}|}^2 &= 
     {10 \over 3 \pi} \kappa_{10}^4 \omega^4 (p_1 \cdot p_2)^2  \cr
    $\ell = 3$:\ \ \ & \overline{|{\cal M}|}^2 &= 
     {5 \over 2 \pi^{3/2}} \kappa_{10}^5 \omega^6 (p_1 \cdot p_2)^2 \ .  \cr
  }}}
 As before, $p_1$ and $p_2$ are the momenta of the outgoing gauge
bosons.  The gauge bosons are identical particles, as are the $\ell$
outgoing scalars, whose momenta will be labeled $p_3$, $\ldots$,
$p_{\ell+2}$.  To avoid over-counting in the integral over phase
space, we must include a factor of $1/(2 \cdot \ell!)$ in the
expression for the cross-section.  The cross-section also includes an
explicit factor of $1/(2 \omega)$ from the normalization of the
incoming dilaton.  Altogether,
  \eqn\CrossF{
   \sigma_{\rm D}^\ell = {N^{\ell+2} \over 2 \cdot \ell!} {1 \over 2 \omega}
    \int {d^3 p_1 \over (2 \pi)^3 2 \omega_1} \cdots
     {d^3 p_{\ell+2} \over (2 \pi)^3 2 \omega_{\ell+2}} \,
     (2 \pi)^4 \delta^4 \big( q - {\textstyle \sum\limits_i} p_i \big) 
    \  \overline{|\M|}^2 \ .
  }
 The momentum integrations take the following form:
  \eqn\IlDef{
   I_\ell = \int {d^3 p_1 \over 2 \omega_1} \cdots
       {d^3 p_{\ell+2} \over 2 \omega_{\ell+2}} \,
     \  \delta^4 \big( q - {\textstyle \sum\limits_i} p_i \big) \ 
       (p_1 \cdot p_2)^2 
    = {3 \pi^{\ell+1} \over 2^{\ell+1}}
       {\omega^{2 \ell + 4} \over (\ell+2)! (\ell+3)!} \ .
  }
 The quickest way to establish \IlDef\ is to Fourier transform to
position space.  

The final results for the first few $\ell$ are
  \eqn\FewSigmas{\eqalign{
   \sigma^{\ell=0}_{\rm D-brane} &= \sigma^{\ell=0}_{\rm \ class.}  \cr
   \sigma^{\ell=1}_{\rm D-brane} &= \sigma^{\ell=1}_{\rm \ class.}  \cr
   \sigma^{\ell=2}_{\rm D-brane} &= 
    \tf{9}{5} \sigma^{\ell=2}_{\rm \ class.}  \cr
   \sigma^{\ell=3}_{\rm D-brane} &= 
    \tf{24}{5} \sigma^{\ell=3}_{\rm \ class.} \ .  \cr
  }}
 If the $\ell=2$ and $\ell=3$ D-brane cross-sections had been smaller
than the corresponding 
classical results, one might have wondered if the
argument around \YoungT\ might possibly be invalidated by some
peculiar interaction.  But any such additional interaction could only
increase the D-brane cross-section, making the disagreement even
worse.

It is worth noting that nowhere in the literature have higher partial
waves been compared with complete success between D-brane models and
General Relativity.  Agreement up to numerical factors was obtained in
\mastI\ for effective string models of four- and five-dimensional
black holes, and independently in \gunp\ for the 
five-dimensional case; but in
the absence of a well articulated prescription for coupling the
effective string to the bulk fields it is difficult to tell much about
numerical coefficients.  

The agreement of the $\ell=1$ cross-section certainly encourages us to
believe that the D-brane description is capable of handling partial
waves correctly.  On the level of effective field theory this may seem
peculiar because, in this low-energy description, the D-brane has no
thickness.  How can an object with no extent in transverse dimensions
absorb particles with angular momentum?  Power counting and group
theory alone seem to dictate the answer \LthPartial, up to
normalization.  

Fortunately, with threebranes the string theory prescription for the
couplings to external fields is more directly accessible than for
other solitonic models of black holes.  While it does indeed appear
that $\ell>1$ partial waves present a test which the DBI action,
supplemented by the prescription of \IK\ to obtain the normalization
of \LthPartial\ via a Taylor expansion, fails to pass, we expect that
a proper string theoretic treatment will once again yield agreement.
For $\ell=1$, a full-fledged disk amplitude computation verifies both
the form of \LthPartial\ and the normalization shown.  The $\ell=1$
disk amplitude, which involves one bulk insertion and three boundary
operators, is easy to deal with because the insertion of one scalar
vertex operator on the boundary simply generates a space-time
translation.  When more than one such operators are present, one
encounters singularities in their mutual collisions that need a
careful treatment. The necessary calculations look quite complicated:
for example, computing the full $\ell=2$ amplitude would be equivalent
via the prescription of \gm\ to computing a six point type~I disk
amplitude. They nevertheless seem
highly worthwhile as a means to refine our understanding of the 
world-volume action.

%%%%%%%%%%%%%%%%%%%%%%%%%%%%%%%%%%%%%%%%%%%%%%%%%
\subsec{Absorption of Higher Spin Particles}
%%%%%%%%%%%%%%%%%%%%%%%%%%%%%%%%%%%%%%%%%%%%%%%%%%%%

In the previous subsections we dealt exclusively with scalar
particles.  Of course, in string theory one also encounters particles
of higher spin, such as the gravitons. The necessary string theory
calculations for them are of the same order of complexity as for
scalars. The comparison with classical gravity is complicated by the
fact that it is difficult in general to decouple the necessary
equations. This problem has been studied to some extent for $d=4$
black holes (see \teukI\ and references therein). Nothing is known,
however, about the $d=7$ black holes relevant to this
discussion. While we hope to come back to the classical aspect of this
problem in the future, below we give a brief discussion of the
effective action results.

In \IK\ absorption of gravitons polarized
transversely to the threebrane was studied with the result that the
cross-section has the same value \SigmaGrav\ as the scalar cross-section.
This is suggestive of a residual supersymmetry relation between
particles of different spin: while the supersymmetry is broken by the 
incident energy, it may still constrain the low-energy limit
of the cross-section. In this section we consider particles that
are vectors from the $d=7$ point of view and find that their absorption
cross-section is $2/3$ times that of the minimally coupled scalar.

One example of a state which gives rise to a vector particle upon
dimensional reduction to $d=7$ is 
obtained from the $B_{MN}$ (see (A.4)).
The spatial and time components of the vector potential are
given by $(B_{\alpha i}, B_{\alpha 0})$, where $\alpha$ is one of the
longitudinal directions (say, $\alpha=x$).
We will choose the gauge $B_{\alpha 0}=0$.

In the world-volume action, the leading term responsible for absorption is
\eqn\beaction{-T_3
\int d^4 x\  {1\over 4}(F_{\alpha\beta}+ \hat B_{\alpha\beta})^2 \ .
}
For the mixed polarization that we are interested in,
$$ \hat B_{\alpha\beta}= B_{\alpha i} \partial_\beta X^i- 
B_{\beta i} \partial_\alpha X^i \ . 
$$
Upon rescaling the world-volume fields by $\sqrt T_3$
and generalizing to $U(N)$, the coupling
is given by
$$ - \int d^4 x\ B_{\alpha i}\tr (  \partial_\beta X^i F^{\alpha \beta}) 
\ .$$
Choosing one particular component, e.g.,
$\a=x, i=7$, and  using the properly normalized field, 
$B_{x7}/(\sqrt 2 \kappa_{10})$ (cf. (A.4)), we get the amplitude
$$ {\cal M} = \sqrt 2 \kappa_{10}
p_1^\beta (p_{2x} \epsilon_\beta - p_{2\beta} \epsilon_1)
\ .$$
Summing over polarizations, we find that
\eqn\vectorm{ \overline{ |{\cal M}|}^2 = 2\kappa_{10}^2 
{\omega^4\over 4} ( 1- n_x^2 )
\ .}
Is there an additional contribution of the same order due to the
world-volume fermions? If we use the prescription of 
\refs{\ceder,\jhs} to supersymmetrize the action \teea, then
we find that the answer is negative: the leading coupling
to the Majorana-Weyl fermion $\Psi$ is via a dimension 6 operator,
$$
\sim B_{\alpha j}\partial_\beta X^j
\bar \Psi \Gamma_{[\alpha} \partial_{\beta]} \Psi
\ .$$
Thus, it appears that the direction dependence of the
absorption process does not cancel out for the $d=7$ vectors.
Since $\vev{n_x^2}=1/3$, we find, multiplying \vectorm\
by the number of final states and
the phase space factor, $1/(16\pi \omega)$, that
the total vector cross-section is given by 
$2/3$ times the cross-section of a minimally coupled
scalar, \SigmaGrav.

It would be interesting to confirm the absence of a leading fermionic
contribution using the string amplitude calculations of \HK.
Perhaps the mixing of perturbations around the 3-brane,
discussed in the Appendix, can explain the factor of $2/3$.

%%%%%%%%%%%%%%%%%%%%%%%%%%%%%%%%%%%%%%%%%%%%%%%%%%%%%%%%%%%%%%%%%
\newsec{Comparing absorption cross-sections for M-branes}
%%%%%%%%%%%%%%%%%%%%%%%%%%%%%%%%%%%%%%%%%%%%%%%%%%%%%%%%%%
A remarkable aspect of the agreement between 
the string theoretic and 
the classical
results for threebranes is that it holds exactly 
for any value of  $N$, including
$N=1$. As explained in the introduction, the classical geometry should
be trusted only in the limit $g_{\rm str} N \rightarrow \infty$.
Thus, the agreement of the absorption cross-sections for $N=1$
suggests that our calculations are valid even in the limit
$g_{\rm str} \rightarrow \infty$, provided that $g_{\rm str}
\alpha'^2 \omega^4$ is kept small. The supersymmetric 
non-renormalization theorems are probably at work here, insuring that 
there are no string loop corrections.

In this section we would like to ask whether the exact agreement
between the absorption cross-sections is also found for the twobranes
and fivebranes of M-theory. These branes have much in common with the
self-dual threebrane of type IIB theory: their entropies scale with
the temperature in agreement with the scaling for a gas of massless
fields on the world-volume \kt, while the scaling of their absorption
cross-sections with $\omega$ agrees with estimates from the
world-volume effective theory \IK. While the effective actions for one
twobrane \bst\ and one fivebrane \ffff\ are now known in some detail,
their generalizations to $N>1$ remain somewhat obscure.\foot{It is
believed, for instance, that $N>1$ coincident fivebranes are described
by non-trivial conformally invariant theories in $5+1$
dimensions. Comparisons with classical gravity of the kind made in
\refs{\kt,\IK} and here are among the ways of learning more about this
theory.}  In this section we compare the cross-sections for $N=1$ and
find that, in contrast to the threebranes, there is no exact agreement
in the normalizations. This is probably due to the fact that M-theory
has no parameter like $g_{\rm str}$ that can be dialed to make the
classical solution reliable.

First we discuss absorption of longitudinally polarized
gravitons by an twobrane. The massless fields in the effective action 
are 8 scalars and 8 Majorana fermions.
The longitudinal
graviton couples to the energy momentum tensor on the
world-volume, $T_{\alpha\beta}$.
The terms in the effective action necessary to describe the absorption of
$h_{xy}$ are ($i= 3, \ldots, 10$;\ $I=1, \ldots, 8$)
\eqn\action{\eqalign{& S_2= T_2
\int d^3 x\  
\bigg [ -\tf{1}{2}  \partial_{\alpha} X^i
\partial^{\alpha} X^i  
+ \tf{i}{2} \bar \psi^I \gamma^\alpha\partial_\alpha
\psi^I\cr & +  \sqrt 2 \kappa_{11} 
h_{xy} \big (\partial_x X^i \partial_y X^i 
- \tf{i}{4}  \bar \psi^I
(\gamma_x \partial_y + \gamma_y \partial_x) \psi^I \big ) 
\bigg ]
\ , }
}
where $h_{xy}$ is the canonically normalized field which enters the
$D=11$ space-time action as (cf. \lgr)
$$ - {1\over 2} \int d^{11} x\ \partial_M h_{xy} \partial^M h_{xy} 
\ .$$

The absorption cross-section is found using the Feynman rules in
a way analogous to the threebrane calculation of section 3.
For the 8 scalars, we find that the matrix element squared 
(with all the relevant factors included) is
$$ {\kappa_{11}^2 \omega^4 \over 2 } 4 n_x^2 n_y^2
\ ,$$
where $\vec n$ is the unit vector in the direction of one of the
outgoing particles.
For the 8 Majorana fermions, the corresponding object 
summed over the final polarizations is
$$ {\kappa_{11}^2 \omega^4 \over 2 } (n_x^2- n_y^2 )^2 \ . 
$$
Adding them up, we find that the dependence on direction cancels out,
just as in the threebrane case. The sum must be multiplied by
the phase space factor 
$$
{1\over 2\omega} {1\over 2\pi} \int {d^2 p_1\over 2 \omega_1} 
\int {d^2 p_2\over 2 \omega_2} 
\delta^2 (\vec p_1 + \vec p_2) \delta (\omega_1+ \omega_2- \omega)=
{1\over 8 \omega^2}
\ ,$$ 
so that the total cross-section is
\eqn\total{ \sigma_{2{\rm \ abs}} = {\kappa_{11}^2 \omega^2 \over 16 } 
\ .}
This does not agree with the classical result for $N$ set to 1,
$$ \sigma_{2{\rm \ class.}}= {2\pi^4\over 3} \omega^2 R^9= {2\sqrt 2\over 3\pi}
\kappa_{11}^2 \omega^2 N^{3/2}
\ ,$$
where we have used the twobrane charge quantization to express $R^9$
in terms of $N$.
Notice that even the power of $\pi$ does not match.
This situation is reminiscent of the
discrepancy in the near-extremal entropy where the relative factor was 
a transcendental number involving $\zeta(3)$\  \kt. 

Now we show that, just as for the threebrane, the transversely
polarized gravitons have the same absorption cross-section as
the longitudinally polarized gravitons.
The coupling of $h_{67}$ to scalars is given by
\eqn\couu{- T_2\int d^3 x\ \sqrt 2 \kappa_{11} h_{67} 
\partial_{\alpha} X^6 \partial^{\alpha} X^7  
\ ,}
while pairs of fermions are not produced because the coupling 
to them vanishes on shell.
The matrix element squared is $\kappa_{11}^2 \omega^4/2$.
Multiplying this by the phase space factor, we again find the 
cross-section \total.

Now we turn to the fivebrane.
The massless fields on the fivebrane form a tensor
multiplet consisting of 5 scalars, 2 Weyl fermions and the
  antisymmetric tensor $\B_{\alpha\beta}$  with anti-selfdual 
strength \refs{\modes}.
The transverse gravitons are again the easier case because they produce
pairs of scalars only (the coupling to 2 fermions vanishes on shell).
The necessary coupling of $h_{67}$ is  similar to \couu, 
$-T_5 \int d^6 x\ \sqrt 2 \kappa_{11} h_{67} 
\partial_{\alpha} X^6 \partial^{\alpha} X^7  . 
$
The matrix element squared is $\kappa_{11}^2 \omega^4/2$,
while the phase space factor is now
\eqn\phase{ {1\over 2\omega}
{1\over (2\pi)^4} \int {d^5 p_1\over 2 \omega_1} 
\int {d^5 p_2\over 2 \omega_2} 
\delta^5 (\vec p_1 + \vec p_2) \delta (\omega_1+ \omega_2- \omega)=
{\omega\over 2^7 \cdot 3 \pi^2}
\ ,}
so that the absorption cross-section is
\eqn\fivecross{\sigma_{5{\rm \ 
class.}}= {\kappa_{11}^2 \omega^5\over 2^8 \cdot 3 \pi^2}
\ .}

To discuss the absorption of longitudinally polarized 
gravitons, $h_{xy}$, we need the action ($i=6, \ldots, 10$; \ 
$I=1, 2$)
\eqn\fiveaction{\eqalign{& S= T_5
\int d^6 x\  
\bigg [ - \ \tf{1}{2} \partial_{\alpha} X^i
\partial^{\alpha} X^i - \tf{1}{12} \H_{\alpha\beta\gamma}^2 
+i \bar \psi^I \gamma^\alpha\partial_\alpha
\psi^I\cr &   + \sqrt 2 \kappa_{11} h_{xy} 
\big (\partial_x X^i \partial_y X^i 
+  \tf{1}{2} \H^-_{x\beta\gamma} \H_y^{- \beta\gamma}
-  \tf{i}{2} \bar \psi^I
(\gamma_x \partial_y + \gamma_y \partial_x) \psi^I \big)
\bigg ]
\ . }
}
To describe interaction of the anti-selfdual antisymmetric tensor
with external field we  follow the covariant approach of 
\alvw\  using the standard `unconstrained' propagator for $\B_{\a\b}$
and replacing $\H$ by  its anti-selfdual part 
$\H^-=\ha(\H- \td \H)$ in the vertices. 

For the 5 scalars, we find that the matrix element squared 
(with all the relevant factors included) is
\eqn\scalar{
{\kappa_{11}^2 \omega^4 \over 2 } {5\over 2} n_x^2 n_y^2
\ .}
For the 2 Weyl fermions, the corresponding object 
summed over the final polarizations is\foot{
It is interesting to observe that, if we consider an 
${\cal N}=1$ multiplet
consisting of 1 Weyl fermion and 4 scalars, then the `tensor term'
$n_x^2 n_y^2$ cancels out in the direction dependence. 
The same cancellation occurs for the threebrane 
(both for the ${\cal N}=1$ vector
multiplet and for the ${\cal N}=1$ hypermultiplet).}
\eqn\weyl{
{\kappa_{11}^2 \omega^4 \over 2 } (n_x^2+ n_y^2 - 4 n_x^2 n_y^2 )
\ .}
Finally,
the contribution  of the anti-selfdual gauge field
turns out to be equal to that  of the
usual, unconstrained 
$\H_{\a\b\g}$  divided by 2.
The matrix element squared is, therefore,\foot{Since the propagator of 
$\B_{\a\b}$  is taken to be non-chiral, it is sufficient to do
the  replacement $\H\H \to \H^-\H^-$ in only 
one of the two  stress tensor factors  in $|{\cal M}|^2$.
The relevant part of 
$\H^-\H^-$ is  $\four (\H \H + \td \H \td \H)$ which is equal to
$ \ha \H \H$ for off-diagonal components of the stress tensor.}
\eqn\anti{|{\cal M}|^2= {1\over 2} 2 \kappa_{11}^2 
{1\over 4} \sum \ [\H^{(1)}_{x\beta\gamma} \H_y^{(2)\beta\gamma}
+ \H^{(2)}_{x\beta\gamma} \H_y^{(1)\beta\gamma}]
\H^{(1)*}_{x\alpha\delta} \H_y^{(2)\alpha\delta *}
\ , }
where we 
have also included $1/2$ because the outgoing particles are identical.
Sums over polarizations are to be performed with
$$ \sum \epsilon_{\alpha\beta}  \epsilon^*_{\gamma\delta}=
\eta_{\alpha\gamma} \eta_{\beta\delta}- \eta_{\alpha\delta}
\eta_{\beta\gamma}
\ .$$
The entire calculation is lengthy, but the end result
is simple,\foot{
This answer passes also a number of heuristic checks. 
For instance, for the ${\cal N}=1$ multiplet
including the gauge field, 1 scalar and 1 Weyl fermion, 
the $n_x^2 n_y^2$ terms again cancels out.} 
\eqn\endi{ \overline{ |{\cal M}|}^2= {\kappa_{11}^2 \omega^4 \over 2 } 
\left (1- n_x^2- n_y^2 + {3\over 2} n_x^2 n_y^2\right ) 
\ .}

Adding up the contributions of the entire tensor multiplet, we find
that all the direction-dependent terms cancel out, just as they did
for the threebrane and the twobrane.
Multiplying by the phase space factor \phase, we find that 
the total cross-section for the longitudinally
polarized gravitons is again given
by \fivecross. This turns out to be
a factor of 4 smaller than the classical result,
\eqn\ende{
 \sigma_{5{\rm \ class.}}= {2\pi^3\over 3} \omega^5 R^9= 
{\kappa_{11}^2 N^3 \omega^5\over 2^6 \cdot 3 \pi^2}
\ ,  } 
evaluated for $N=1$, i.e. 
$\sigma_{5{\rm \ abs}} = \four \sigma^{(N=1)}_{5{\rm \ class.}}$.
This discrepancy is relatively minor and is of a kind that could easily
be produced by a calculational error. However, having checked
our calculations a number of times, we believe that the 
factor of 4 discrepancy is real.

A conclusion that we may draw from this section is that, although the
single M-brane cross-sections scale with the energy in the same way
as the classical cross-sections, the normalizations do not agree.
The fivebrane comes much closer to agreement than the twobrane,
which may be connected to
the fact that the fivebrane supergravity solution is
completely non-singular. For $N=1$, however, the curvature of the 
solution is of order of the 11-dimensional Planck scale.
Obviously, the 11-dimensional supergravity is at best a low-energy
approximation to M-theory. The M-theory effective action should contain
higher-derivative terms weighted by powers of $\kappa_{11}$, by
analogy with the $\alpha'$ and $g_{\rm str}$  
expansions of the string effective action.
Thus, for $N=1$, the classical solution may undergo corrections of order
one which we believe to be the source of the discrepancy.
For large $N$, however, we expect the M-theory cross-section to agree
exactly with the classical cross-section.
We hope that these considerations 
will serve as a useful guide in constructing the world
volume theory of $N$ coincident fivebranes.

%%%%%%%%%%%%%%%%%%%%%%%%%%%%%%
\newsec{Conclusions}
%%%%%%%%%%%%%%%%%%%%%%%%%%%%%%%%
In this paper we have provided new evidence, furthering the earlier
results of \IK, that there exists exact agreement between the 
classical and the
D-brane descriptions of the self-dual threebrane of type
IIB theory. The specific comparisons that we have carried out involve
probing an extremal threebrane with low-energy massless quanta incident
from the outside. As argued in \IK\ and here, the 
great advantage of the
threebrane is that both the perturbative
string theory and the classical supergravity
calculations are under control and yield expansions in the same
dimensionless expansion parameter 
$(\omega R)^4 \sim N g_{\rm str} \alpha'^2 \omega^4$,
which may be kept small.

While the low-energy physics of the threebrane is described by a ${\cal N}=4$
SYM theory, probing it from the extra dimensions provides a new point
of view and allows for a variety of interesting calculations.
One example is the absorption of a graviton polarized parallel
to the brane which, as discussed in section 3,
couples to the energy-momentum tensor in
3+1 dimensions. Because of the existence
of the transverse momentum, the kinematics for this process is that of
a decay of a massive spin-two particle into a pair of
massless world-volume modes. Summing over all the states
in the ${\cal N}=4$ multiplet we find that the rate is completely
isotropic, which is undoubtedly related to the conformal invariance
of the theory. We should emphasize, however, that we are exploring
the properties of this theory away from the BPS limit; 
therefore, they are not determined by supersymmetry alone.
For this reason, we find it remarkable 
that the net absorption cross-section
has the same value as that of the dilaton and the RR scalar, 
and which agrees with the cross-section found in 
classical supergravity.

We have also carried out similar explorations of the physics
of M-branes. These studies are hampered, to a large extent, by 
insufficient understanding of the multiple coincident branes of
M-theory. For a single twobrane, a formal application of classical 
gravity gives an answer which is off by a 
factor of $32\sqrt 2/(3\pi)$,
while for a single fivebrane -- off only by a factor of 4.\foot{Curiously, 
both mismatch factors are of the same order, as $32\sqrt 2/(3\pi) 
\approx 4.8$.}
For the fivebrane the discrepancy is relatively minor which, we are
tempted to speculate, is due to the fact that its classical geometry
is completely non-singular, just like that of the threebrane.
We should keep in mind, however, that for singly charged branes
the quantum effects of M-theory are expected to be
important, and the classical reasoning should not be trusted.

While in this paper we have probed 3-branes  with massless 
particles,  we may contemplate another interesting use of the 3-brane:
it may  be used as a probe by itself \mdlate.
For example, $N$ coincident 3-branes may be probed by
a 3-brane parallel to them. This situation is described by
a $3+1$ dimensional ${\cal N}=4$ supersymmetric 
$U(N+1)$ gauge theory, with gauge symmetry broken
to $U(N)\times U(1)$. A more complicated theory will arise if
a 3-brane is used to probe a stringy
$d=4$ black hole.
As was  shown in \refs{\KT,\bl}, 
the $d=4$
extremal black holes with regular horizons 
(which are parametrised by 4 charges \CY)  
can be represented by 1/8 supersymmetric configurations of four intersecting 
3-branes wrapped over a  6-torus.
 Following
\refs{\mepr}  one can find an action for a classical 3-brane 
probe moving in this geometry. If the probe is oriented parallel
to one of the four source 3-branes, the resulting moduli space 
metric is 
$$ds^2_6=  H_1 H_2 H_3  (dr^2 + r^2 d\Omega_2^2)  + 
H_1 dy^2_1  +  H_2 dy^2_2 +  H_3 dy^2_3
\ , \ \ \ \ \ H_i =1 + {R_i\ov r}
\  , $$
where $y_i$ are toroidal coordinates transverse to the probe.
In contrast to the case of the `5-brane+string+momentum'
configuration   describing  $d=5$  regular  extremal 
black holes with 3 charges \ATT\ 
where one  finds   \dpl\ that the non-compact part of the moduli space 
 metric is multiplied by the product of two  harmonic functions, here 
we obtain  the product of  {\it three}
 harmonic functions. This is, however,
exactly what is needed to get the same near-horizon 
($r\to 0$) behaviour as found in \dpl, 
i.e. that the 3-dimensional non-compact
part of the moduli space metric becomes flat: 
$ds^2_3 \to  R_1R_2R_3 (d\r^2 +  \r^2   d\Omega_2^2) $, \
 $\r\equiv {r}^{-1/2} \to \infty$.
This close similarity 
between a $d=5$ black hole probed by a string  
and a $d=4$ black hole probed by a 3-brane
strongly suggests that 
one can obtain important information 
about these black holes  by studying the corresponding 
3-brane world-volume theory.
This theory in the presence of intersecting  branes is necessarily 
more intricate than the simpler, parallel brane case
discussed in the main part of this paper.

Clearly, there is much work to be done before we claim a complete
understanding of the remarkable duality between the D-brane and the
classical descriptions of the threebrane.  The
complexities that affect the higher-spin classical equations remain to
be disentangled.  The numerical discrepancies for $\ell>1$ partial
waves suggest an incomplete understanding of the low-energy effective
action.  As we have argued in section 3, a direct string calculation
is necessary to check normalizations.  And finally, we need to push
our methods away from extremality where the physics is necessarily
more complicated, involving thermal field theory in 3+1 dimensions. We
hope that detailed insight into such a theory will help explain the
specific factors appearing in the near-extremal entropy \GKP.

We feel that further efforts in the directions mentioned above are
worthwhile because they offer a promise of building a theory
of certain black holes
in terms of a manifestly unitary theory -- perturbative string
theory. 

\newsec{Acknowledgements}

We are grateful to V.~Balasubramanian, C.G.~Callan, and A.~Hashimoto
for useful discussions.  This work 
was supported in part by DOE grant DE-FG02-91ER40671,
the NSF Presidential Young Investigator Award PHY-9157482, and the
James S.{} McDonnell Foundation grant No.{} 91-48. 
A.A.T. also acknowledges  
the  support of PPARC and 
 the European
Commission TMR programme ERBFMRX-CT96-0045.

\vfill\eject

\appendix{A}{Comments on mixed perturbations}

In this Appendix we discuss the problem of mixing between different
perturbations around the self-dual threebrane.
We may attempt to classify
various perturbations according to the power of the
harmonic  function
$H(r)$  which appears in  front of their kinetic term.
For example, the off-diagonal graviton perturbations (the $d=7$ 
Kaluza-Klein vectors $A^a_\m = g^{ab}h_{b\m}$)  
have the following kinetic term  
\eqn\kkv{  S_{vect.grav.} 
= {1\ov 2\k_{10}^2}  \int d^{10} x  \sqrt{-g_{10}}  
\big(
 - { \textstyle{1\ov 4}} g_{ab} F^{a\m\n} F^b_{\m\n} + ... \big) }
$$
= -{1\ov 8\k_{10}^2}  \int d^{10} x\   H^{-1} (r) \big[
 F^a_{ij } F^a_{ij}  -  2 H(r)   F^a_{0j } F^a_{0j}  \big]    + ...  \ . $$
Because of the 
extra factor of $H\inv$ in the
kinetic term  one expects that 
 their low-energy absorption rate is suppressed 
 compared to that for
the minimally coupled massless  scalars in \scal,\ligr\  (cf. \CGKT).
However,
Eq. \kkv\ is not the complete action 
for the  vector perturbations  $\delta A^a_\m$, as they  `mix' with
  fluctuations of  $C_4$. 
The components of $C_4$ which vanish in the 
3-brane  background are $C_{\m\n ab}$
(antisymmetric 2-tensors in $d=7$) 
and $C_{\m\n\l a}$ (antisymmetric 3-tensors dual to vectors in $d=7$). 
These two types of components are related  by the self-duality condition, 
$$F_{ijk ab} \sim \ep_{abc}  \ep_{ijklmn} F_{0lmn c}\ ,\qquad
F_{0jk ab} \sim  H^{-1} \ep_{abc} \ep_{jklmni} F_{lmni c}\ .
$$
The action for the quadratic fluctuations  of the first field
has the same structure as the scalar actions \scal,\ligr\
($F_{\m\n\l ab} = { 3}  \del_{[\m} C_{\n\l] bc}$)\foot{The action 
for $C_{\m\n\l a}$  has an extra factor of   $H^{-1}$
in front of $[ F_{ijkl a} F_{ijkla} - 4 H(r)  F_{0jkl a} F_{0jkla}]$, and 
its absorption  should be  suppressed at low energies. This  
  will follow also from world-volume considerations  discussed 
below:  while  $C_{ij ab}$ couples to  the `marginal' $(dX^i)^2$  operator, 
$C_{ijk a}$  couples to the $(dX^i)^3$-term which
produces a smaller cross-section at low energies.} 
\eqn\ftce{  S_{tens.} 
= {1\ov 2\k_{10}^2}  \int d^{10} x  \sqrt{-g_{11}} \big(
- { \textstyle{1 \ov 48 }}   F_{\m\n\l ab} F^{\m\n\l ab}\big)}
$$
=\  -
{1\ov 96\k_{10}^2}  \int d^{10} x   
 [ F_{ijk ab} F_{ijkab} - 3 H(r)  F_{0jk ab} F_{0jkab}] \ . $$
The coupling  between  $C_{\m\n  ab }$   and 
the graviton perturbation $h_{\m a}= \delta A^a_\m$ 
originates from  the  $F^2_5$  term in \efc\  
(or  from  the self-duality condition  implying 
$F_{0ij ab} \sim \ep_{abc} \del_i H\inv   h_{jc}$)  
and has the following structure, 
\eqn\mix{\sim  \ep_{abc} H\inv  \del_i H F_{0ij ab} h_{jc} \ .  }
A similar pattern is found for 
  the structure of the 2-nd rank antisymmetric tensor
perturbations  in \efc.
The   fields $B_{ab}, C_{ab}$ (which become
scalars in $d=7$)  and $B_{\m a}, C_{\m a}$
(which become vectors in $d=7$) 
are  non-trivially coupled pairwise ($B_{ab}$ with $C_{\m a}$
and $C_{ab}$ with $B_{\m a}$)
 via the  WZ-term  in \efec\
and  the `cross-term' in $F^2_5$. 
Both contributions to the coupling
 have the same form because of the self-duality of the background value of 
$\del C_4$. The  kinetic
 part of their action is  (cf. \scal,\kkv,\ftce)
  $$ S_{vect.} 
= {1\ov 2\k_{10}^2}  \int d^{10} x  \bigg(
- { \textstyle{1 \ov 4 }}  H (r) \big[ (\del_i B_{ab})^2  
- H(r) (\del_0 B_{ab})^2 + 
(\del_i C_{ab})^2  - H(r) (\del_0 C_{ab})^2\big] 
$$ 
\eqn\feee{ - \ { \textstyle{   }}  
\big[ (\del_{[i} B_{j] a})^2  - 2H(r) (\del_{[0} B_{i]a})^2
+ (\del_{[i} C_{j] a})^2  - 2H(r) (\del_{[0} C_{i]a})^2 \big]\bigg)\ , 
}
and the mixing terms are
\eqn\coup{
\sim \ep_{abc} \del_i H \big( C_{ab} \del_{[0} B_{i]c} - B_{ab} \del_{[0} C_{i]c} \big)
\ . } 
While  the scalars $B_{ab},C_{ab}$ have extra  factors of $H$ 
in front of their kinetic terms, 
%(enhacing their low-energy absorption)
the vectors $B_{\m a}, C_{\m a}$ have the standard
kinetic terms  as in  \scal.
We hope that the mixing which affects these $d=7$ vectors
is capable of reproducing the peculiar factor of $2/3$ 
multiplying their absorption cross-sections, found in section 3.4
using the effective field theory on world-volume.

%\appendix{B}{ }

%Here we offer some 
Analogous  observations  can be made on 
the problem of mixing between different
perturbations around the M-brane solutions.
The Kaluza-Klein vector field  has the action 
of the same form as  \kkv\ in the type IIB case, i.e.
with the   factor  $H\inv$ in front of the kinetic term.
In the twobrane (fivebrane) case it 
 mixes (cf. \mix) with the electric 
(magnetic)  component of the field strength of 
 $C_{\m\n a}$, i.e. with the antisymmetric 
2-tensors in $d=11-p$ dimensions.
For the twobrane, the tensor $C_{\m\n\r}$ 
 has kinetic term $H\inv F_4^2$
and extra coupling  term $\ep_{9} \del H\inv F_4 F_4$, while
$C_{\m ab}$, which has a non-trivial background value,
 should  mix with   `transverse  gravitons' $\delta g_{\m\n}$.
A similar mixing is expected for $C_{\m\n\l}$
in the fivebrane case.

For the  fivebrane    
there is also  another pair of `mixed'  fields (analogous  to the 
system  in \feee,\coup):
$C_{abc}$ (scalars in $d=6$)
and $C_{\m ab}$ (vectors in $d=6$).
Their kinetic terms are  as  in \feee,
 $$ S_{tens.} 
= {1\ov 2\k_{11}^2}  \int d^{11} x  \bigg(
- { \textstyle{1 \ov 12 }}  H (r) \big[ 
(\del_i C_{abc})^2  - H(r) (\del_0 C_{abc})^2\big] 
$$ 
\eqn\eee
{ -\  { \textstyle{3 \ov 4 }}  \big[ (\del_{[i} C_{jk] a})^2  - 2H(r) (\del_{[0} C_{ik]a})^2  \big]\bigg)\ ,  
}
and the  mixing term  originating from the WZ-term in \eleva\
 is\foot{Upon reduction to $d=6$ 
the relevant interaction term may be written as 
$ W^{\m\n} F_{ab\m\n} S^{ab}$, where 
$W^{\m\n} \sim  \ep^{\m\n\k\l\r\s} (\del C)_{\k\l\r\s}$
is the  dual vector field strength, 
 $F_{ab\m\n}$ is the field strength of the vectors $C_{ab\m}$
and $S^{ab} \sim \ep^{abcde} C_{cde}$ are the scalars.
In the  fivebrane background  the 
vector  field  strength  $W_{\m\n}$  has a non-vanishing
electric component which produces mixing between 
the electric component of $F_{ab\m\n}$ and the scalars.}
\eqn\miii{
\ep^{\m\n\l\r\k\s}  \ep^{abcde}
(\del C)_{\m\n\l\r} (\del C)_{\k\s ab} C_{cde}
\sim \ep_{abcde}\del_i H (\del C)_{0i ab} C_{cde}\ .}
The effect of this mixing on the absorption cross-sections
remains to be explored.

\listrefs
\bye